\documentclass[12pt]{article}

\newtheorem{Proposition}{Proposition}
\newtheorem{Lemma}{Lemma}

\newcommand{\roha}{\mbox{${\rho_H}$}}
\newcommand {\cA}{\mbox{${\mathcal A}$}}

\newcommand {\cH}{\mbox{${\mathcal H}_x$}}
\newcommand {\cHH}{\mbox{${\mathcal H}$}}
\newcommand {\cB}{\mbox{${\mathcal B}$}}
\newcommand {\cM}{\mbox{${\mathcal M}$}}
\newcommand {\RR}{\mbox{${\mathbf{R} }$}}

\begin{document}

\title{Geometry of Non-Hausdorff Spaces and Its Significance for Physics}

\author{Michael Heller\thanks{Correspondence address: ul.
Powsta\'nc\'ow Warszawy 13/94, 33-110 Tarn\'ow, Poland. E-mail:
mheller@wsd.tarnow.pl} \\
Vatican Observatory, V-00120 Vatican City State \\
Copernicus Center for Interdisciplinary Studies \\
ul. S{\l}awkowska 17, 31-016 Cracow, Poland \\
\and Leszek Pysiak and Wies{\l}aw Sasin \\
Department of Mathematics and Information Science, \\ Warsaw
University of Technology \\
Plac Politechniki 1, 00-661 Warsaw, Poland \\
Copernicus Center for Interdisciplinary Studies \\
ul. S{\l}awkowska 17, 31-016  Cracow, Poland \\}

\date{\today}

\maketitle

\begin{abstract}
Hausdorff relation, topologically identifying points in a given space, belongs to elementary tools of modern mathematics. We show that if subtle enough mathematical methods are used to analyze this relation, the conclusions may be far-reaching and illuminating. Examples of  situations in which the Hausdorff relation is of the total type, i.e., when it identifies all points of the considered space, are the space of Penrose tilings and space-times of some cosmological models with strong curvature singularities. With every Hausdorff relation a groupoid can be associated, and a convolutive algebra defined on it allows one to analyze the space that otherwise would remain intractable. The regular representation of this algebra in a bundle of Hilbert spaces leads to a von Neumann algebra of random operators. In this way, a probabilistic description (in a generalized sense) naturally takes over when the concept of point looses its meaning. In this situation counterparts of the position and momentum operators can be defined, and they satisfy a commutation relation which, in the suitable limiting case, reproduces the Heisenberg indeterminacy relation. It should be emphasized that this is neither an additional assumption nor an effect of a quantization process, but simply the consequence of a purely geometric analysis.

\end{abstract}

\section{Introduction}
Hausdorff relation, topologically identifying points in a given space, belongs to elementary concepts of modern mathematics. In this paper we show that it can lead to far-reaching consequences if it is ``resolved'' with the help of suitably sensitive tools. Non-Hausdorff spaces, often regarded as a technical nuisance, sometimes produce a global disaster. This happens when large domains of the considered space, or even the entire space, topologically collapse to a single point (in the latter case, we say that the Hausdorff relation is of the total type). The sign of such a disaster is that only functional algebras consisting of constant functions can be defined on these domains or spaces. Examples of such situations are many: the space of Penrose tilings \cite{Penrose}, the leaf space of foliations \cite{Connes}, space-times of some cosmological models with strong curvature singularities \cite{Bosshardt, Johnson}. As it is well known, with every equivalence relation a groupoid can be associated allowing one to employ various techniques typical for groupoids \cite{Paterson}. It turns out that these techniques, when applied to the groupoid associated with the Hausdorff relation, lead to interesting conclusions. This is less surprising if we notice that the Hausdorff relation is strictly connected with the  individuality of points in a given space (Lemma 1 below) or, in other words, with the very nature of this space.

In Carlo Rovelli's opinion, the main challenge of the present search for quantum gravity theory is ``understanding what is a background independent quantum field theory'' \cite[p. 8]{Rovelli}. The meaning of the term itself ``background independent theory'' is not very clear, but the most radical versions of its understanding claim that we should get rid of points altogether. This is exactly what happens when the Hausdorff relation is of the total type. But then the traditional approach in terms of coordinates and atlases gives place to a ``spectral'' approach in terms of noncommutative algebras and their representations. The so called regular representation (in the bundle of Hilbert spaces) of a noncommutative algebra on the Hausdorff groupoid  leads to a von Neumann algebra (called von Neumann algebra of the groupoid) the elements of which are random operators. It seems natural that loosing information about points results in a probabilistic type of reasoning, albeit in a generalized sense (probability measures are given here by normal, faithful and normalized to unity states on the von Neumann algebra). As the consequence of this, we are able to formulate the commutation relation between two operators which, in the limit to the Hausdorff case, reproduces the known commutation relation between position and momentum operators. It should be emphasized that this result is purely geometric and does not depend of any particular physical assumptions.

The composition of our material runs as follows. In Sect. 2, we formulate Hausdorff relation in an algebraic language and, in Sect. 3, we give the construction of the Hausdorff groupoid and a noncommutative algebra on it. In Sect. 4, we present the von Neumann algebra of random operators associated with the Hausdorff groupoid. In Sect. 5, we consider a (finite) sequence of algebras (and the Hausdorff groupoids corresponding the them) such that the first algebra in the sequence corresponds to the situation in which all points are non-Hausdorff separated from each other, and the last algebra in the sequence to the space satisfying the Hausdorff axiom. We call this the noncommutative deformation of the latter algebra. In Sect. 6, we compute the commutator of the generalized position and momentum operators.  Finally, in Sect. 7, we make some interpretative remarks.

\section{Hausdorff Equivalence Relation}
Let us consider a family $C$ of real valued functions on a set $M$. We endow $M$ with the weakest topology, denoted $\tau_C$, in which the functions of $C$ are continuous. A function $f$ defined on $U \subset M$ is said to be a local $C$-function if, for every $x \in U$, there exist a neighborhood $V$ of $x$ in the topological space $(U, \tau_U)$, where $\tau_U$ is the topology in $U$ induced by $\tau_C $, and a function $g \in C$ such that $f|V = g|V$. Let us denote by $C_U$ the set of all such local $C$-functions. We have $C \subset C_M$. If $C = C_M$, the family $C$ is said to be closed with respect to localization. If, for any $n \in \mathbf{N}$ and every function $\omega \in C^{\infty }(\RR^n)$, we have $f_1, \ldots , f_n \in C$ implies $\omega \circ (f_1, \ldots , f_n) \in C$ then the family $C$ is said to be closed with respect to superposition with Euclidean functions.
If the family $C$ is closed with respect to localization and closed with respect to superposition with Euclidean functions, the family $C$ is said to be a differential structure on $M$, and the pair $(M, C)$ a differential space. Any function belonging to $C$ is said to be smooth (in the sense of differential spaces). If $C = C^{\infty }(M)$, the differential space is a differential manifold (for the theory of differential spaces see \cite{Structured}). In the following we shall assume that $(M, C)$ is a differential manifold or a differential space.

We define the Hausdorff relation $\roha \subset M \times M$ in the following way $x\roha y \Leftrightarrow \forall_{f\in C} f(x) = f(y)$, $x,y \in M$. It is clearly an equivalence relation.

Let $\rho \subset M \times M$ be any equivalence relation. We define two families of functions. First,
\[
C/\rho = \{\bar{f}: M/ \rho \rightarrow \RR : \bar{f} \circ \pi_{\rho }\}
\]
where $\pi_{\rho }:M \rightarrow m/\rho $ is the canonical projection. This family is the largest differential structure on $M/\rho $ in which $\pi_{\rho }$ is smooth (in the sense of the theory of differential spaces). Second,
\[
C_{\rho } = \{f \in C: \forall_{x,y \in M} \; f(x) = f(y) \}.
\]
Functions belonging to $C_{\rho }$ are called $\rho $-consistent.

There exists the isomorphism of linear algebras $\Phi: C/\rho \rightarrow C_{\rho }$ given by $\Phi(\bar{f}) = \bar{f} \circ \pi_{\rho }$ for $\bar{f} \in C/\rho $.

\begin{Lemma}
The algebra $C_{\rho_H}$ of $\rho_H$-consistent functions coincides with the algebra $C$, i.e., every function of $C$ is \roha -consistent.
\end{Lemma}

\noindent
\textit{Proof.} On the one hand, $x \roha y \Rightarrow f(x) = f(y) \Rightarrow f \in C_{\rho_H}$, therefore $C \subset C_{\rho_H}$. On the other hand, $C_{\rho_H} \subset C$ as its subalgebra. $\diamond $

This lemma says that the Hausdorff relation \roha \ is intimately related to the differential structure $C$. It is, in a sense, responsible for the individualization of points in a given space. We clearly have the differential space $(M/\rho_H, C_{\rho_H})$ the points of which are
$[x] = \{y \in M: \forall_{f \in C} f(y) = f(x)\}$.

\section{Hausdorff Groupoid}
As it is well known, each equivalence relation gives rise to a groupoid \cite{Connes}. It turns out that, in the case of the Hausdorff equivalence relation, the corresponding groupoid proves to be very useful in the study of even strongly non-Hausdorff spaces. We call it Hausdorff groupoid and denote by $\Gamma $. We now briefly describe its construction.

We identify $\Gamma $ with the graph of the Hausdorff relation
\[
\roha = \{(x,y) \in M \times M: \forall_{f \in C} f(x) = f(y)\}.
\]
Let $(x,y), \, (y,z) \in \Gamma $, then the multiplication is given by $(x,y) \circ (y,z) = (x,z)$, the inverse by $(x,y)^{-1} = (y,x)$, and the identity by $(x,x)$. The source and target mappings are given by $d(x,y) = x$ and  $r(x,y) = y$, respectively. The following sets are defined:
the set of elements beginning at $x$, $\Gamma_x = \{(x,y): x \roha y\} = d^{-1}(x)$, the set of elements ending at $y$,
$\Gamma^y = \{(x,y): x \roha y \} = r^{-1}(y)$, and the isotropy group $\Gamma_x^x = \Gamma_x \cap \Gamma^x =\{(x,x): x \in M\}$. In general the groupoid $\Gamma $ is not transitive. 

Let us first assume that the relation \roha \ is of the finite type, i.e., that the number of points identified under \roha \ is denumerable. We define the algebra $\cA = C_c^{\infty }(\Gamma)$ of smooth, complex valued, compactly supported functions on $\Gamma $ with the convolution 
\[
(f * g)(x,y) = \sum_{x \roha z \roha y} f(x,z)g(z,y),
\]
$f,g \in \cA $ , as multiplication. We also define the involution by $f^*(x,y) = \overline{(y,x)}$.

If a differential space $(M,C)$ is Hausdorff then, for every $x \in M$, $[x]_{\roha } = \{x\}, \; (f*g)(x,x) = f(x,x)g(x,x)$ and $\Gamma = \Gamma ^{(0)} = \{(x,x): x\in M\}$ is diffeomorphic with $M$. In consequence, the algebra \cA \ is isomorphic with $C$. 

Let us define the Hilbert space $\cH =L^2(\Gamma^x)$ and the bundle of Hilbert spaces $\cHH = (\cH )_{x \in M}$. Then we have the regular representation of the algebra \cA \ in \cH , $\pi_x: \cA \rightarrow \mathcal{B}(\cH)$, given by
\[
(\pi_x(a)\psi )(x,y) = \sum_{x \roha z}a(x,z)\psi (z, y),
\]
for $a \in \cA, \, \psi \in \cH $. If the support is compact, the sum is finite. We thus have
\[
\cA \ni a \mapsto \pi(a) = (\pi_x(a))_{x \in M} := r_a.
\]

\section{Von Neumann Algebra Associated with Hausdorff Relation}
In sections 4 -- 5, we consider the case when $M$ is a differential manifold of a finite dimension and the Hausdorff relation \roha \ is of the total type, i.e., $[x]_{\roha } = M$ for every $x \in M$.

We define the algebra $\cA = C_c^{\infty }(M \times M)$ and the convolution
\[
(a * b)((x,y) = \int_M a(x,z) b(z,y) d\lambda (z),
\]
for $a,b \in \cA $, where $d\lambda $ is the usual manifold measure. Since the functions of \cA \ are of compact support, the integral is well defined.

We define the representation $\pi_x: \cA \rightarrow \cB(\cHH )$ of the algebra \cA \ in the Hilbert space $\cH = L^2(\Gamma^x)$ by
\[
(\pi_x(a))(z,x) = \int_M a(z,y) \psi (y,z) d\lambda(y),
\]
for $a \in \cA , \psi  \in \cHH $, and consider the isomorphic image $\cM_0$ of \cA \ by $\pi := (\pi_x)_{x \in M}$
\[
\cM_0 = \pi(\cA ) = \{(\pi_x (a))_{x \in M}, a \in \cA \}.
\]

The algebra $\cM_0 $ can be completed to the von Neumann algebra \cM , $\cM_0'' = \cM $, where $\cM_0'$ is a commutant of $\cM_0 $ in the Hilbert space $ \int_{\oplus } \cH d\lambda(x)$. This von Neumann algebra is called von Neumann algebra of the groupoid $\Gamma $ \cite{Paterson}. 

\begin{Proposition}
Von Neumann algebra \cM \ is an algebra of random operators.
\end{Proposition}

\noindent
\textit{Proof.} 
Random operator acting on a Hilbert bundle $H = (H_x)_{x \in X}$ is a function
\[
A: X \ni x \mapsto A_x \in \cB(H_x)
\]
such that
\begin{enumerate}
\item
If $\{\psi_i\}_{i=1}^{\infty }$ is a field of bases in $H_x, x \in X$, i.e., if $\{\psi_i(x)\}_{i=1}^{\infty }$ is a base in $H_x$, then the function
\[
X \ni \rightarrow (A_x \psi_i(x), \psi_j(x))_x \in \RR ,
\]
$i,j = 1,2, \ldots $, is measurable,
\item
the family of operators $A_x, x \in X$, is essentially bounded in the operator norm, i.e.,
$\mathrm{ess\, sup} ||A_x|| < \infty$.
\end{enumerate}
It is straightforward to check that the operators $r_a = \{\pi_x(a)\}_{x \in M}$, for every $a \in \cA $, satisfy these conditions. $\diamond $

The concept of randomness is an element of the probability panorama. We shall now reconstruct the probabilistic environment for this concept. Let us notice that $\cH = L^2(\Gamma^x)$ is isomorphic with $L^2(M)$ and let us consider the functional
\[
\Phi (r) = \int_M \mathrm{tr}_x(\hat{\rho r}) d\lambda (x),
\]
for $r \in \cM $, where the density function on $M$, $\hat{\rho }: M \rightarrow \cB(\cH )$, is such that
\begin{enumerate}
\item
$\hat{\rho }$ is trace class, i.e., for every $x \in M$, $\mathrm{tr}_x\hat{\rho }(x)  < \infty $,
\item
$\mathrm{tr}_x\hat{\rho } \in L^1(M)$ is an integrable function,
\item
for every $x \in M, \, \hat{\rho }(x)$ is a positive operator,
\item
the normalization condition holds $\int_M \mathrm{tr}_x(\hat{\rho }(x)) d\lambda (x) = 1$.
\end{enumerate}

The functional $\Phi(r)$ with the above conditions is called a normal state on the von Neumann algebra \cM . If additionally $\hat{\rho } $ is an injection, the state $\Phi (r)$ is called faithful. 

In noncommutative mathematics, not necessarily commutative algebras are regarded as leading to the concept of generalized probability space. More strictly, 
the noncommutative probability space is defined to be a pair $({\cal M},\varphi )$  where ${\cal M}$ is a von Neumann
algebra and $\varphi$ a faithful and normal state on ${\cal M}$ \cite{Biane}. Noncommutative probability theory is quickly developing branch of modern mathematics \cite{Cuculescu}. In contrast to the commutative case, the noncommutative case exhibits the great richness of possibilities. In a noncommutative regime, probabilistic properties of a physical system can strongly depend on the state in which this system finds itself.
\par

\section{Noncommutative Deformation of a Manifold}
So far we considered two extreme cases: (1) when the relation $\roha \subset M \times M$ was of the finite type, i.e., when the number of points identified under \roha \ was denumerable; (2) when the Hausdorff relation $\roha \subset M \times M$ was of the total type, i.e., when all points of $M$ were identified under \roha . When the Hausdorff relation \roha \ is not of the total type, the Hausdorff groupoid $\Gamma $ is a disjoint sum of groupoids $\Gamma_{[x]}$ such that $\Gamma_{[x]} = [x] \times [x]$, where $[x ]$ is the equivalence class of $x$, i.e., every component groupoid $\Gamma_{[x]}$ is of the total type.

As an example let us consider the following situation. Let $M$ be a measurable subset of $\RR^n$, and let us consider the differential space $(M,C)$ with the differential structure on $M$ induced from $\RR^n$, i.e.,  $C = C^{\infty }(\RR^n)_M$. On the strength of the embedding theorem, $M$ can be assumed to be a differential manifold. Let now $\Gamma = \rho \subset M \times M$ where $\rho $ is any equivalence relation. We say that $\rho $ is of measurable type if the equivalence class $[x]$ is measurable for every $x \in M$. We define the algebra $\cA = C^{\infty }_{fc}(\Gamma )$ of smooth functions on $\Gamma $ which are compact along fibres of $\Gamma $ (``fibre compact'') with the convolutive multiplication
\[
(a*b)(x,y) = \int_{[x]}a(x,z)b(z,y)d\mu(z),
\]
$a,b \in \cA , x,y \in \Gamma $. We assume that $[x]$ is of measurable type.

\begin{Proposition}
If $\tau_{c/\rho }$ is $T_1$ then $\rho $ is of measurable type.
\end{Proposition}

\noindent \textit{Proof.} If $\tau_{C/\rho }$ is $T_1$ then the sets $\{[x]\}$, for any $x \in M$, are closed in $(M/\rho , \tau_{C/\rho })$. Consequently, the equivalence class is closed in $(M, \tau_C)$ since $[x] =\pi_{\rho }^{-1}([x])$ and the mapping $\pi_{\rho }$ is continuous. Therefore, the subset of the form $[x] \subset M, x \in M$, as a closed set, is measurable. $\diamond $

In the following, we shall consider the projections $\pi_i|M \rightarrow \RR$ given by $(\pi_i|M)(x_1, \ldots , x_n) = x_i$, and proceed in the subsequent steps.

\textit{Step $0$.} Let $C_0 = \mathrm{Gen}\{\mathbf{1}\}_M \cong \RR$ be a differential structure on $M$, and let $\rho_0 = M \times M$ be defined by $x \rho_0 y \Leftrightarrow \forall_{\alpha \in C_0} \alpha (x) = \alpha (y)$. This relation is of the total type and everything is as in Sections 4 and 5.

\textit{Step $k, k = 1, \dots , n-1$.} The differential structure $C_k$ is generated by projections, $C_k = \mathrm{Gen}\{\pi_1, \ldots , \pi_k\}$, and $x \rho_k y\Leftrightarrow \pi_i(x) = \pi_i(y)$ for $i = 1, \ldots , k$. The Hausdorff groupoid is
\[
\Gamma_k = \{(x,y) \in M \times M: \pi_i(x) = \pi_i(y), \, i = 1, \ldots , k\}
\] \[
 = \bigcup_{(\pi_1, \ldots , \pi_k)(x) \in (\pi_1, \ldots , \pi_k)(M)} [x] \times [x].
\]
Therefore, the set $[x] \in M/\rho_k$ is closed, and $\{[x]\} = \{\pi_1, \ldots , \pi_k \}(t), \, t \in (\pi_1, \ldots , \pi_k)(M)$. It follows that $\rho $ is of measurable type. We thus can write
\[
(a *_k b)(x,y) = \int_{[x]} a(x,z)b(z,u) d\mu (z)
\]
for $a,b \in \cA_k = C_{fc}^{\infty }(\Gamma_k )$.

\textit{Step } $n$. We obviously have $x \rho y \Leftrightarrow \pi_i(x) = \pi_i(y), \, i = 1, \ldots , n \Leftrightarrow x = y$, and $\Gamma_n = \bigcup_{x \in M} \{x\} \times \{x\} \bigcup_{x \in M}\{x,x\} = \Delta $, and $[x] = \{x \}$. Therefore, $\cA_n = C_{fc}^{\infty }(M)$ and $a *_n b = a \cdot b$.

Summarizing, we have the following sequence of groupoids
\[
\Gamma_0 \stackrel{\iota_0}{\supset} \Gamma_1 \stackrel{\iota_1}{\supset} \ldots \stackrel{\iota_{k-1}}{\supset } \Gamma_k \stackrel{\iota_k}{\supset }\ldots \stackrel{\iota_{n-1}}{\supset } \Gamma_n ,
\]
and the following sequence of the corresponding algebras
\[
\cA_0 \stackrel{\iota_0^*}{\rightarrow } \cA_1 \stackrel{\iota_1^*}{\rightarrow } \ldots \stackrel{\iota_{k-1}^*}{\rightarrow }\cA_k \stackrel{\iota_k^*}{\rightarrow }\ldots \stackrel{\iota_{n-1}^*}{\rightarrow } \cA_n 
\]
where $\iota_s^* $, $s = 0, \ldots , n$, are homomorphisms of algebras. In this sense, we can speak about the noncommutative deformation of the algebra $\cA_n $.

\section{Position and Momentum}
Let us consider the derivation $P: C^{\infty }(M) \rightarrow C^{\infty }(M)$ of the algebra $C^{\infty }(M)$ (of course, with the
pointwise multiplication). We shall also write $(P(x)f)(x) =
(Pf)(x)$ for $f \in C^{\infty }(M)$. We lift the derivation $P$ to
the product $M\times M$ in two ways:
\[
\bar{P}_{hor}(x,y) = (\iota_y)_{*x}P(x)
\]
and
\[
\bar{P}_{ver}(x,y) = (\iota_x)_{*y}P(y)
\]
where $\iota_y(x) = (x,y)$, $\iota_x(y) = (x,y)$. Clearly, $\bar{P}_{hor}, \bar{P}_{ver} \in \mathrm{Der}(C^{\infty
}(M \times M))$. Remembering that $\cA \subset C^{\infty }(M\times
M)$, we make the restrictions
\[
\bar{P}_{hor}|\cA : C_c^{\infty }(M\times M) \rightarrow C_c^{\infty
}(M\times M),
\]
\[
\bar{P}_{ver}|\cA : C_c^{\infty }(M\times M) \rightarrow C_c^{\infty
}(M\times M),
\]
and perform the symmetrization
\[
\bar{P}(x,y) = \bar{P}_{hor}(x,y) + \bar{P}_{ver}(x,y).
\]

We now check that $\bar{P}$ is a generalized derivation of the
convolution algebra $C_c^{\infty }(M\times M)$. First, we observe that
\[
(\bar{P}(a * b)) (x,y)= \int_M \bar{P}(x,y) (a(x,z)\cdot
b(z,y))d\lambda (z).
\]
Next, we check that $\bar{P}$ satisfies the generalized Leibniz rule,
i.e.,
\[
\bar{P}(a * b) = \bar{P}_{hor}(a) * b + a * \bar{P}_{ver}(b).
\]
In checking this we should take into account that $\bar{P}$
decomposes into the horizontal component and the vertical component
in the following way
\[
\bar{P}(a(\cdot , z) \cdot b(z, \cdot )) = \bar{P}_{hor}(a(\cdot ,
z)) \cdot b(z, \cdot ) + a(\cdot, z) \cdot \bar{P}_{ver}(b(z, \cdot
)).
\]
Thus we have
\[
(\bar{P}(a * b)) (x,y)= \int_M \bar{P}_{hor} (a(x,z))\cdot b(z,y) +
a(x,z)\cdot\bar{P}_{ver}(b(z,y)) d\lambda (z)=
\]
\[
= (\bar{P}_{hor}(a) * b)(x,y) + (a * \bar{P}_{ver}(b))(x,y).
\]

Let us notice that the algebra $\cA = C_c^{\infty }(M\times M)$ is a $Z$-bimodule, where $Z = C^{\infty }(\Delta )$, $\Delta = \Gamma^{(0)}$ being the diagonal of $M \times M$. And clearly $Z$ is isomorphic with $C^{\infty }(M)$. We consider the left action of $Z$ on \cA, $\alpha : Z \times \cA \rightarrow \cA $, defined in the following way
\[
\alpha (f, a) (x,y) = f(x,x) \cdot a(x,y)
\]
for $f \in Z, a \in \cA $. Let us denote $\alpha(f,a) = (Q(f))a$. We
have the following commutation relation
\[
[\bar{P}, Q(f)] = Q(Pf).
\]
Indeed, since $\bar{P}$ satisfies the usual Leibniz condition for
pointwise multiplication we have
\[
 [\bar{P}, Q(f)](a)(x,y) =\bar{P}(Q(f)(a))(x,y)-
(Q(f)\bar{P}(a))(x,y) =
\]
\[
=\bar{P}(f(x,x)\cdot a(x,y)) - f(x,x)\cdot \bar{P}(a(x,y))=
\]
\[
=(Pf)(x,x)\cdot a(x,y)+ f(x,x)\cdot \bar{P}(a(x,y)) -f(x,x)\cdot
\bar{P}(a(x,y))=
\]
\[
=(Q(Pf)a)(x,y).
\]
 In the $n$th step of the previous section we have $\cA =
C_{fc}^{\infty }(M)$ and, if $M = \RR^n$, we obviously have $[\pi_i,
P_i] = \mathbf{1}$ which means that the above equation is indeed a
generalization of the usual Heisenberg commutation relation.

\section{Concluding Remarks}
At first sight it is surprising that a simple (non-)Hausdorff property could have such far-reaching consequences. However, on the second thought this seems to be more understandable. Lemma 1 of Sect. 2 states that the family of functions consistent with the Hausdorff relation on a given manifold (or on a differential space) coincides with the differential structure of this manifold, and it is precisely the differential structure that fixes both topological and differential properties of the manifold. Therefore, we are here at the very basis of the concept of space. 

Usually groupoids are regarded as generalizations of groups (``groups with many units''), but they can also be regarded as generalized equivalence relations. ``From this point of view, a groupoid over $B$ tells us not only which elements of $B$ are equivalent to one another, but it also parametrizes the different ways in which two elements can be equivalent`` \cite{Weinstein}. Alan Weinstein compares the groupoid structure to that of gauge theories. In these theories space-time points are equipped with internal group symmetries in such a way that, when they are composed into a gauge theory, some of their irreducible representations correspond to elementary particles. ``But gauge groups and gauge transformations are applicable only because all the points of space-time are alike'' \cite{Weinstein}. In the case of the Hausdorff groupoid, we are also taking into account ``different ways'' in which any two points are equivalent, and this can be regarded as a kind of internal symmetries with which the points (and not any elementary particles) are equipped. The nontrivial consequence of this fact is that, after defining a convolutive algebra on a given groupoid and representing it on a bundle of Hilbert spaces, random properties of the construction emerge. It is interesting that in this situation counterparts of the position and momentum operators can be defined, and they satisfy a commutation relation which, in the suitable limiting case, reproduces the Heisenberg indeterminacy relation well known in quantum mechanics. It seems reasonable that if we loose the identity of points, the only possibility to predict something must be via a probabilistic kind of reasoning. 

Of course, not always a suitable groupoid is a simple Hausdorff groupoid, explored in the present paper. In more advanced cases some more complex types of groupoids should be used (we have chosen to deal with the simplest possible case to show, as clearly as possible, arcana of its working). But the strategy is always the same: when the Hausdorff property is broken down, groupoid symmetries are surfacing and the noncommutative regime takes over. 

We should realize that the nonlocal and probabilistic character (noncommutation relation of generalized position and momentum operators included) are neither additional assumptions nor effects of a quantization process, but conclusions deduced from purely mathematical premises. It is therefore natural to propose a hypothesis that the nonlocal and probabilistic character of quantum mechanics are but ``hereditary'' properties which have somehow survived the ``phase transition'' from the noncommutative regime to quantum mechanical physics. If so, these two properties (nonlocality and probabilistic character) must be present at the fundamental level (albeit in a generalized form). To be sure, this conclusion does not preclude looking for quantum gravity theory in various directions.

\end{document}